\documentclass[preprintnumbers,showpacs]{revtex4}
\usepackage{graphicx}
\usepackage{color}

\def\bc{\begin{center}}
\def\ec{\end{center}}

\def\beq{\begin{equation}}
\def\eeq{\end{equation}}

\begin{document}

\title{On  Binding Energy of Trions in Bulk Materials}
\author{Igor Filikhin$^{1}$, Roman Ya. Kezerashvili$^{2,3}$,  Branislav Vlahovic$^{1}$}

\affiliation{\mbox{$^{1}$ Mathematics and Physics Department, North Carolina Central University, }\\
Durham, NC 27707, USA \\
\mbox{$^{2}$ Physics Department, New York City College
of Technology, The City University of New York,} \\
Brooklyn, NY 11201, USA \\
\mbox{$^{3}$The Graduate School and University Center, The
City University of New York,} \\
New York, NY 10016, USA}

\begin{abstract}
We study the negatively $T^{-}$ and positively $T^{+}$ charged trions in bulk materials 
in the effective mass approximation within the framework of a potential model.
The binding energies of trions in various semiconductors are calculated by 
employing Faddeev equation in configuration space. Results of calculations 
of the binding energies for $T^{-}$ are consistent with previous computational studies 
and are in reasonable agreement with experimental measurements, while the $T^{+}$ is 
unbound for all considered cases. The mechanism of formation of the binding energy of trions 
is analysed by comparing contributions of a mass-polarization term related to kinetic energy operators 
and a term related to the Coulomb repulsion of identical particles.
\end{abstract}


\maketitle

Excitonic effects in semiconductors are determined
by the exciton binding energy and electron-hole interaction and play a critical role in
optoelectronic devices \cite{Bim}. Charged exciton complexes such as negative ($T^-$) and
positive ($T^+$) trions are formed when a single exciton is correlated with an additional electron in a 
conduction band or hole in a valence band, respectively, 
has been proposed by Lampert \cite{L}. In the meantime $T^-$ and $T^+$ trions have been the subject
of intense studies in the last two decades, both experimentally and theoretically. Their observation in
bulk semiconductors has been hampered due to their rather small binding energies and became a
challenging task. 
Trions were first 
observed in quantum wells \cite{QW1} in 1993
and shortly thereafter in GaAs-AlGaAs quantum wells \cite{QW2,QW3,QW4}. Trions were
were predicted and found in  the photoluminescence and
absorption spectra of various optically excited semiconductors,
especially in quantum dots \cite{BonitzCD,El17}, quantum wells
\cite{Bim,Es,Es1} and carbon nanotubes \cite{CNT1,CNT2}. Mott-Wannier trions in
two- and three-dimensional (2D and 3D) semiconductors can be described
by the solutions of the three-body Schr\"odinger equation after modelling
the crystal by effective electron and hole masses and a dielectric
constant. There are stochastic variational calculations, and
studies by using density functional theory, variational quantum Monte
Carlo method, and the diffusion Monte Carlo approach \cite{Varga1999,DifMC,RPV2000}.
Calculations have shown that the
binding energy of a trion is strongly enhanced in two-dimensional
structures due to the trion's larger spatial extent. Trions have been observed in 2D 
transition-metal
dichalcogenide (TMDC) semiconductors \cite{MoS23Heinz,MoSe21 Ross,WSe2 Jones,WSe2Wang,MoSe2Singh,Liu,
ShangBiexiton,WS2Plechinger,ZhangMS2,WS2ZHU2015}. Until now several approaches have
been proposed for evaluations of the binding energies of a trion in two-dimensional transition metal
dichalcogenides. Initial work on trion binding energies in TMDCs employed variational wave functions
\cite{Reichman2013}, and more recently using the time-dependent density-matrix functional
theory, the fractional dimensional space approach, the stochastic variational method
with explicitly correlated Gaussian basis, method of hyperspherical harmonics, and
quantum Monte Carlo methods, such as the diffusion Monte Carlo and the path integral
Monte Carlo \cite{Reichman2013,Thilagam2014,TimeDepFankTheor,DenFuncTheoryPIMC,Saxena,Ganchev,BerkelbachDifMonteCarlo,
VargaPRB2016,SpinkDifMC,KezFBS2017,Falko,Falko2}. 
Let us note that trions are studied in anisotropic two-dimensional materials such as phosphorene and arsenene \cite{C}, and are predicted to have remarkably high binding energies.
Though much progress has been made, intrinsic excitonic states of 2D and 3D trions are
still highly debated in theory, particularly related to the 
binding energies for negatively and positively charged trions which
thirsts for experimental determination. In this letter we address this issue.

Because trions are intrinsically three-particle objects,
common calculation methods are not always adequate to describe their
behavior and a more rigorous level of theory must be
employed. In the present work we study the $T^-$ and $T^+$ trions within the Faddeev
formalism \cite{MF85}, -- the most rigorous approach 
for investigating a three-body system.
In the case of a trion one deals with a three-body system $AAB$ with two
identical particles. We perform ground-state calculations for 
a positively and negatively charged 
trion in the effective mass approximation within the framework
of a nonrelativistic potential model using the method of Faddeev equations in
configuration space \cite{MF85}. 
This approach gives new insights to the problem, because the Faddeev equations are 
the most general equations for description of a non-relativistic three-particle 
system within the potential approach and use as inputs only 
masses of particles and pairwise inter-particle interaction. 
There are no any fitting parameters in our approach. 
In the case of trions in bulk the inter-particle interaction 
is described by the Coulomb potential and the electron and hole masses 
can be obtained by different {\it ab initio} methods: many-body $G_0W_0$ and $GW$, 
density functional theory and the local density approximation. Therefore, one can 
understand what kind other quantum effects, which are not included in 
the potential model, should be considered for an adequate description of trions. 
In our approach,
the trion is a three-particle system $eeh$ ($hhe$) 
consisting of electrons ($e$) and  heavy holes ($h$), with each pair interacting by
the  Coulomb  force. To understand the
origin of the difference of binding energies for  the  mirror systems of charged trions, we solve the Faddeev equations 
for cases when all three particles are interacting via the 
Coulomb potential and when the interaction between two identical
particles is omitted  or is screened and use these solutions to analyze contributions of two terms which 
course the difference of binding energies. The first term is related to the Coulomb repulsion between
two identical particles and the second one is the mass-polarization term (MPT) \cite{H2002} 
related to the kinetic energy operators. 
The latter term can be most clearly introduced by using the Schr\"{o}dinger equation 
in the system of reference 
relative to the non-identical particle. 
 
The Faddeev equations in configuration space
can be written in the form of a system of second order differential 
equations \cite{MF85},
which can be reduced to a simpler form for the case of two identical particles. In this
case the total wave function of the system is decomposed into the sum of the Faddeev
components $U$ and $W$ corresponding to the $(AA)B$ and $(AB)B$ types of
rearrangements: $\Psi =U+W\pm PW$, where $P$ is the permutation operator for two
identical particles. In the latter expression the sign ''$+$'' corresponds
to two identical bosons, while the sign ''$- $'' corresponds to two identical fermions,
respectively. After introducing the set of the Jacobi coordinates for the three particles,
separating the motion of the center-of-mass
one can write the set of Faddeev equations for the relative motion of three particles 
when two of them are identical fermions in the following form \cite{FGS1,KezFil}:
\begin{equation}
\begin{array}{l}
{(H_{0}+V_{AA}-E)U=-V_{AA}(W - PW)}, \\
{(H_{0}+V_{AB}-E)W=-V_{AB}(U - PW)}.
\end{array}
\label{GrindEQ__1_}
\end{equation}%
In Eq. (\ref{GrindEQ__1_}) the Hamiltonian $H_{0}$ is the operator of kinetic energy
written in terms of corresponding Jacobi coordinates, while $V_{AA}$ and $V_{AB}$
are the potentials of the pairwise interactions between the particles.
The pairwise interactions are described by the Coulomb potential
with the dielectric constant related to the considered material.

Let us consider the states of $T^{-}$ and $T^{+}$ trions with the total angular
momentum $L=0$, the momentum of pair $l=0$, and momentum
$\lambda=0$ of the third particle with respect to the center-of-mass of the pair.
Within this condition the pair of electrons (holes) is in a singlet spin state.
The corresponding spin function is asymmetric with respect to the permutation operator $P$,
which provides automatically the asymmetry of the trion wave function $\Psi$: $P\Psi =P(U+W- PW)=-U+PW-W=-\Psi$.

To analyze the origin of the difference of binding energies for the $T^-$ and $T^+$
trions let us follow Ref. \cite{H2002} and write the Schr\"{o}dinger equation for the trion in the system of 
reference 
relative to the non-identical particle: 
\begin{equation}
(-\frac{\hbar^{2}}{2\mu}
\nabla_{r_{A_1}}^{2}-\frac{\hbar^{2}}{2\mu}\nabla_{r_{A_2}}^{2}-\frac{\hbar^{2}}{m_B}\nabla_{r_{A_1}}\nabla_{r_{A_2}}+V_{AB}(r_{A_1})+V_{AB}(r_{A_1})-V_{AA}(r_{A_1}-r_{A_2})-E_3)\Psi(r_{A_1},r_{A_2},r_{A_1}-r_{A_2})=0,
 \label{Sh}
\end{equation}
which is written in a self-explanatory notation. In Eq. (\ref{Sh}) $\mu$ is the reduced
mass of the electron and hole and
$T_{mp}=-\frac{\hbar^{2}}{m_B}\nabla_{r_{A_1}}\nabla_{r_{A_2}}$ is the mass-polarization
term and $E_3$ is the ground state energy of the three particles. In the case $m_B < m_A$ the contribution of the MPT 
can be of the same order as the contribution of the other
two differential operators in Eq. (\ref{Sh}) due to the comparable mass factors of
these operators,  which can be expressed as $1/m_B$. In the case $m_B > m_A$ the
contribution of this term has the factor $1/m_B$, while the mass factors of other
differential operators are of the order of $1/m_A$. When $m_B >> m_A$ the contribution of
the MPT can be ignored. If in Eq. (\ref{Sh}) the MPT 
and the interaction $V_{AA} \equiv V_{AA}(r_{A_1},r_{A_2},r_{A_1}-r_{A_2})$ between two identical 
particles are  neglected one obtains:
\begin{equation}
(-\frac{\hbar^{2}}{2\mu}
\nabla_{r_{A_1}}^{2}-\frac{\hbar^{2}}{2\mu}\nabla_{r_{A_2}}^{2}+V_{AB}(r_{A_1})+V_{AB}(r_{A_1})-E_3(T_{mp}=0,V_{AA}=0))\Psi(r_{A_1},r_{A_2})=0,
 \label{ShR}
\end{equation}
where $E_3(T_{mp}=0,V_{AA}=0)$ is the ground state energy of the three particle system for the 
aforementioned condition. In Eq. (\ref{ShR}) the total wave function can be factorized as
$\Psi(r_{A_1},r_{A_2})=\Phi(r_{A_1})\Phi(r_{A_2}),$ where $\Phi(r_{A_1})$ is
a solution of  two-body Schr\"{o}dinger equation for the $AB$ subsystem,
and that leads to the trivial  solution:   $E_3(T_{mp}=0,V_{AA}=0)=2E_2$, where $E_2$ 
is the ground state energy of two-body subsystem $AB$. Within our consideration 
the bound $AB$ pair is the exciton.  

To evaluate the effect of the MPT, one can neglect the interaction between the identical 
particles, $V_{AA}=0$, in Eq. (\ref{Sh}). 
The evaluation can be written as $\Delta =E_3(T_{mp}=0,V_{AA}=0)-E_3(V_{AA}=0)$,
where  $E_3(V_{AA}=0)$ is the ground state energy of three particle system $AAB$ 
when interaction between the identical particles is neglected. 

Taking into account Eq. (\ref{ShR}), the last expression can be rewritten as
\begin{equation}
\Delta=2E_2-E_3(V_{AA}=0)=B_3(V_{AA}=0)-2B_2\ge 0,
\label{mp}
\end{equation}
where $B_2$ and $B_3(V_{AA}=0)$ are the binding energies of the exciton and three-body $AAB$ system when the interaction between 
two identical particle is omitted, respectively. In the simplest case when $m_B >> m_A$ the contribution of the MPT can be neglected and one has: $B_3(V_{AA}=0,<T_{mp}>=0)=2B_2.$
In consequence, Eq.  (\ref{mp}) is valid for any mass ratio $m_B/m_A$ and can be used
to evaluate the effect of the mass polarization term. The
relation (\ref{mp}) is known in nuclear physics as the mass polarization
effect \cite{H2002,FG2002,FKSV17}. 

Let us introduce the interaction between two identical particles as $\alpha <V_{AA}>$, where
the parameter $\alpha$ controls  the strength of this interaction. 
Substituting this potential in Eq. (\ref{Sh}) and averaging it gives 
the following expression:
\begin{equation}
E_3=-<T_1>-<T_2>+ <V_{AB}>+<V_{AB}>-<T_{mp}>+\alpha <V_{AA}>.
 \label{Sh2_0}
\end{equation}
In Eq. (\ref{Sh2_0}) the matrix elements $<T_1>=<T_2>$ and $<V_{AB}>=<V_{AB}>$ due to the symmetry of the system. 
Thus, from the one hand by solving of the Faddeev equations 
(\ref{GrindEQ__1_}) one can find binding energies $B_3$ for the $T^-$ and $T^+$ trions and 
test the sensitivity of their binding energy to the strength of $\alpha <V_{AA}>$ by 
varying the parameter $\alpha$.  
On the other hand, by solving (\ref{GrindEQ__1_}) 
under the condition $V_{AA}=0$, one can find the binding energies $B_3(V_{AA}=0)$ for 
trions when the interaction between two identical particles is omitted. Such an approach 
allows one to analyze the origin of the binding energy difference for $T^-$ and $T^+$ trions. 
Obviously, $<T_{mp}>$ is related to $\Delta$ as well as $<T_{mp}>\approx \Delta$ when the 
contribution of MPT to $E_3(V_{AA}=0)$ is small.

Let us discuss results of calculations for excitons and $T^{\pm}$ trions binding energies in 
various bulk semiconductors. The binding energies $B_3$ and $B_3(V_{AA}=0)$ for the trions 
were calculated using the the aforementioned Faddeev formalism. We calculate 
the binding 
energy $B_2$ of the exciton in the same semiconductor as well. In our calculations we use the known 
mass ratio ${m_B}/{m_A}$ for various bulk materials and the corresponding 
dielectric constant $\epsilon$. The numerical results of our calculations presented in Table~\ref{t1} 
show that the negative trion is 
always bound, while the positively charged trion has no bound state for all set of parameters. 
To demonstrate that for the particular equal masses the negatively 
and positively charged trions have the same energies and can be 
bound or unbound depending on the value of the dielectric constant 
we perform calculations for trions binding energies in bulk MoS$_2$ which are 
presented in Table \ref{t1}.
The analyses of the results 
for $B_3(V_{AA}=0)$ and binding energy of the exciton $B_2$ shows that the relation  
$B_3(V_{AA}=0)>2B_2$ is always satisfied for  trions and relative contribution of the MPT indicated by 
$\delta=(B_3(V_{AA}=0)-2B_2)/B_3(V_{AA}=0)$. For the positively charged trions, $\delta $ increases when the 
ratio ${m_B}/{m_A}$ decreases.  For the negatively charged trions, the MPT effect is small and  $\delta \sim 0$.

\begin{figure}[t]
    \centering
    \includegraphics[width=3.in]{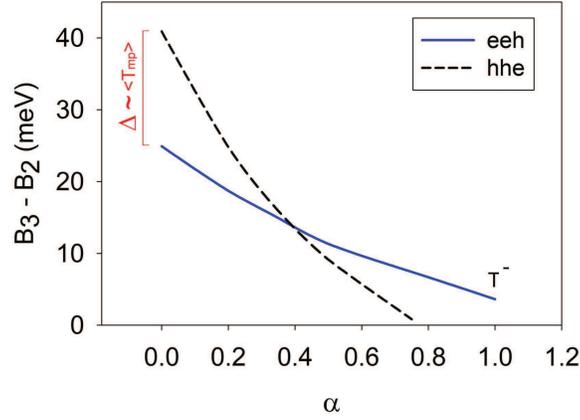}
    \caption{The $B_3-B_2$ for  $eeh$ (solid curve) and $hhe$ (dashed curve)  in InN as a function of the scaling factor 
	$\alpha$  of the Coulomb repulsion between identical particles. The crossing point corresponds to $\alpha=0.41$. The relative value of the mass polarization 
	term $<T_{mp}>$ for $hhe$ is shown by the vertical line segment. }
     \label{2bfigure}
\end{figure}

\begin{figure}
    \centering
       \includegraphics[width=3.8in]{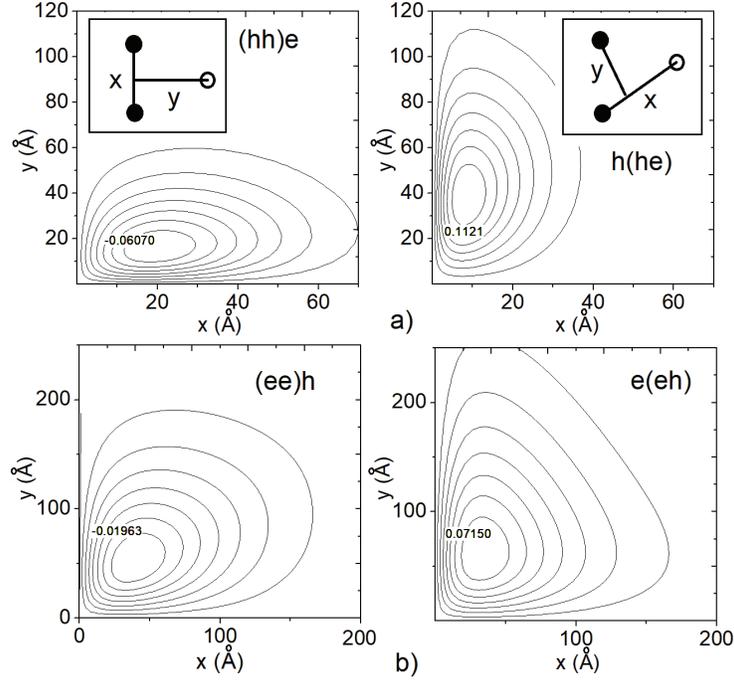}
    \caption{The contour plots of the Faddeev components  U (left) and  W (right) for a) $hhe$ and b) 
	 $eeh$ systems  when the scaling factor for the $ee$ (or $hh$) Coulomb repulsion $\alpha=0.41$ 
	 (see Fig. \ref{2bfigure}). The inserts show the corresponding Jacobi trees and Jacobi 
	 coordinates $x$ and $y$ are measured in {\r A}.}
     \label{3figure}
\end{figure}

\begin{figure}[t]
    \centering
    \includegraphics[width=5.45in]{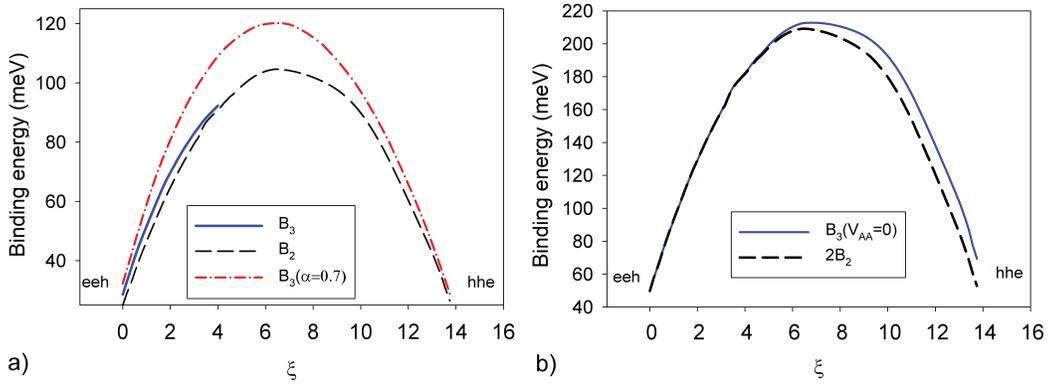}
    \caption{The mass transformation from $eeh$ to $hhe$ in InN. a) The binding energies $B_3$ (solid curve), 
	$B_2$ (dashed curve) and $B_3(\alpha=0.7)$  (dot-dashed curve); and b) $2B_2$  (dashed curve), and $B_3(V_{AA}=0)$ (solid curve), 
	as a function of the mass transformation parameter $\xi$. 
	The parameter $\xi$  is  related to the negative trion, when $\xi$=0, and to the
positive trion, when $\xi$=13.75.}
    \label{1figure}
\end{figure}

Let us mention  that the results 
for $B_2$ confirm the hydrogenic exciton energy 
$B_2=Ry^*=13.61\mu^*/\epsilon_r^2$ (eV)  ($Ry^*$ is the effective Rydberg constant 
with the reduced electron-hole effective mass $\mu^*$) and are in good agreement with 
experimental data  and theoretical calculations \cite{Es1,DWW, RPV2000,AYK,Ak,A,HV2016}.
\begin{table}[ht]
\caption{The charged and neutral exciton binding energy deference $B_T=B_3-B_2$ (the binding energy of trion 
with respect to the exciton binding energy) for different materials. $B_2$ and $B_3$ are the binding energy 
for the trion and exciton, respectively.
  The relative contribution of the mass polarization term $\delta$ is calculated as  $\delta=(B_3(V_{AA}=0)-2B_2)/B_3(V_{AA}=0)$.
 Here $\hbar^2/m_0$=7.6195~eV{\AA}$^2$, $\epsilon$ is  the dielectric constant. All energies are given in meV and 
 the masses are given in units of a free electron mass  $m_0$. 
 }
\centering
\label{t1}      
\begin{tabular}{ccccccccc}
\hline\noalign{\smallskip}
Material: $m_e$, $m_h$, $\epsilon$& ${m_h}/{m_e}$&${m_e}/{m_h}$&Trion &$B_2$ & $B_3$&$B_3(V_{AA}=0)$& $B_T$ &$\delta $\\[3pt]
\hline \noalign{\smallskip}
InN: 0.11, 1.63, 7.5 \cite{DWW}  &14.8 & &$T^-$&24.88 & 28.5 &49.8 & 3.6  & $\sim$0\% \\
                                             & &0.07 &$T^+$&24.88 & --    &65.8  &-- &24\% \\
 \noalign{\smallskip}\hline
GaAs: 0.067, 0.51, 12.9&7.6 &  &$T^-$&4.83 & 5.33 &9.66 & 0.5  & $\sim$0\% \\
                      & &0.13&$T^+$ &4.83 & -- &11.8 &-- &  18\% \\
 \noalign{\smallskip}\hline
ZnSe: 0.16, 0.75, 8.6&4.7&  &$T^-$&24.2 & 26.3 &48.4 & 2.1 & $\sim$0\% \\
        &   & 0.2 &$T^+$ &24.2 & -- &55.7 & -- &13\%\\
       \noalign{\smallskip}\hline
GaN: 0.2, 0.82, 8.9& 4.1 & &$T^-$&27.57 &29.6&55.1 & 2.1  & $\sim$0\% \\
               & &0.24&$T^+$ &27.57& -- &62.4 & --  & 12\% \\
       \noalign{\smallskip}\hline
CdTe: 0.096, 0.35, 10.16& 3.6& &$T^-$&9.91 &10.6 & 19.8 &  0.6  & $\sim$0\% \\
                    & &0.27&$T^+$ &9.91 & -- &22.2 & --  & 11\% \\
       \noalign{\smallskip}\hline
MoS$_2$: 0.45, 0.45, 12.6 \cite{Sa}& 1.0& &$T^-$&19.3 & -- &39.3 & -- & 1.7\% \\
   \ \ \   \ \ \ \ \ \ \ &  & 1.0&$T^+$&19.3 & -- &39.3 & -- & 1.7\% \\
\ \ \ \ \ \ \ \   0.45, 0.45, 10.7 \cite{Ch,Mo}&   1.0 &     & $T^-$  &26.7 & 26.8 & 54.5 &$\sim$ 0.1 & 2.0\% \\
     \ \ \   \ \ \ \ \ \ \ &    &  1.0  & $T^+$  &26.7 & 26.8 & 54.5 &$\sim$ 0.1 & 2.0\% \\
\noalign{\smallskip}\hline
\end{tabular}
\end{table}
Fig. \ref{2bfigure} presents the difference $B_3-B_2$ for 
$AAB$ systems as a function of $\alpha$, which controls 
the strength of interaction between identical particles: $V_{AA}\to \alpha V_{AA}$. The difference 
between the curves at $\alpha =0$ shows the contribution of MPT for the $hhe$ system 
relative to the $eeh$. The slopes of the curves differ significantly. 
This indicates that 
the repulsion between identical particle is much stronger in the $hhe$ system than in $eeh$ because the 
two holes are localized much closer to each other due to their larger effective masses. 
Thus, our hypothetical model with parameter $\alpha$, which controls 
the strength of interaction between identical particles for both trions and effectively represents    
screening caused predominantly by the valence electrons, leads to a weaker 
Coulomb interaction between the identical particles and  hence an increased trion 
binding energy. However, this interaction is stronger for two holes  
because they are localized more closely to each other than electrons, due to the 
larger kinetic energy caused by the heavier effective mass of holes, 
and hence a reduced trion binding 
energy. Therefore, the effect of strong repulsion due to the Coulomb interaction 
takes place. This fact is illustrated in Fig. \ref{3figure} by the contour plots of the Faddeev 
component U and W: the negative trion has more extended distribution 
within about 200$\times$250~{\r A}, than less extended distribution within 
80$\times$120~{\r A} for the positive trion. The analyses of Figs. \ref{2bfigure} 
and \ref{3figure} allows to conclude that both systems are bound with 
the same binging energy  when $\alpha=0.41$, However, the $hhe$ is more compact.
One can make the transformation of $T^{-}$ to $T^{+}$ by replacing of the masses of 
the electron and hole 
and vice versa ($m_e \to m_h$ and $m_h \to m_e$). The parameter $\xi \geq 0$  sets 
this transformation as follows: 
$m^\xi_A= (1+\xi)m_A, m^\xi_B= (1-\xi m_A/m_B)m_B$ and keeps the 
sum of the masses, $m_A+m_B=const$. 
For example, for the effective masses of the electron and hole in InN, when $\xi$=0 
we have $T^{-}$, while for $\xi$=13.75 we have $T^{+}$. The dependence of  
different characteristics for $eeh$ and $hhe$ systems as a function of the parameter $\xi$ 
is presented in Fig. \ref{1figure}.   During the transformation,  the energy $E_3$  follows 
$B_2$,  does not return to the initial value 
defined for $\xi$=0 due to the strong increasing of the Coulomb contribution (see Fig. \ref{1figure}a)).
The contribution of the Coulomb repulsion increases more quickly than 
the $B_2$ when the ratio $m_B/m_A$ 
increases.

The replacement $e \to h$ in the $eeh$ system, when the $AA$ interaction 
is omitted,  leads to  the 
increase of the MPT contribution indicated by the increasing $B_3(V_{AA}=0)$ relatively $2B_2$  (Fig. \ref{1figure}b)). 
Note that, in case of GaAs, the relative increase varies from $0$\% to $18$\% or, in absolute values, 
from 9.66~meV to 11.8~meV. 

The analyses of the results presented in Fig. \ref{1figure} shows that the matrix element of the 
Coulomb repulsion 
$<V_{AA}>$ increases more quickly than the MPT,  $<T_{mp}>$. 
In another words, small variations of the masses (making more compactness of the system) generate larger  
Coulomb repulsion between identical particles. The latter results in the unbound state of the transforming 
$eeh$ system and the final $hhe$ system ($T^{+}$).

One can reduce this Coulomb repulsion artificially using the parameter $\alpha$. 
The results are presented in Fig. \ref{1figure}a). 
The energy of $T^{+}$ is evaluated as 0.7~meV above the $eh$ threshold when $\alpha$=0.7. 
In this case, $E_3$ is larger $B_2$ during the transformation and in the final point. It means also that 
the MPT can compensate for the increased Coulomb repulsion.

\begin{figure}[t]
    \centering
    \includegraphics[width=7in]{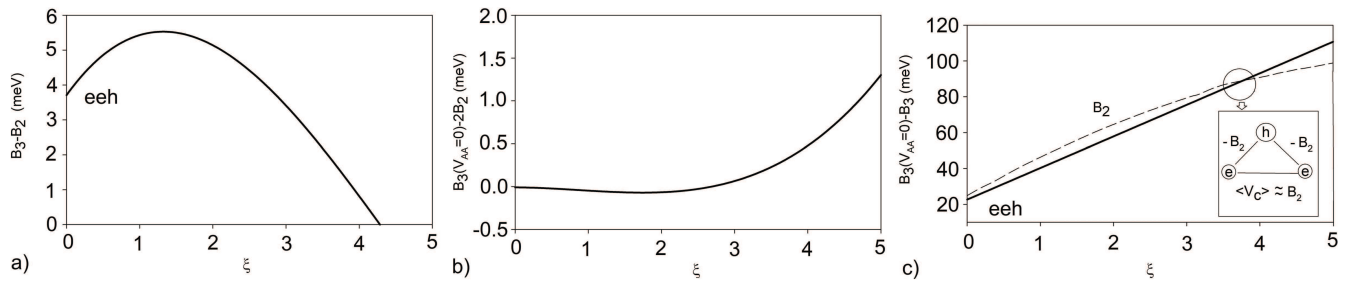}
    \caption{The value of a) $B_3-B_2$, b)  $B_3(V_{AA}=0)-2B_2$, c) $B_3-B_3(V_{AA}=0)$ for  $eeh$  in InN as a function of the 
	mass transformation parameter  $\xi\le 5$ (solid curve). 
     The $B_2$ is shown by the dashed curve. The inset shows schematic distribution for energies in the system. }
     \label{2afigure}
\end{figure}

The Coulomb repulsion of $AA$ particles  can be evaluated as $\Delta B_c(AA)=B_3-B_3(V_{AA}=0)$.
This is shown in Fig. \ref{2afigure}, which presents the dependence of different characteristics for the $eeh$ 
system on the parameter $\xi$=0. One can conclude that, firstly, the binding energy 
of $eeh$ system $B_T=B_3-B_2$,   $B_T/B_2 << 1$  decreases during the mass transformation to the
negative value (unbound state) when  $\xi \sim 4$  (Fig. \ref{2afigure} a)).
One can further conclude, secondly, the 
contribution of the MPT increases with the increment of $\xi$ and hence when the mass 
ratio $m_B/m_A$ increases. However, the MPT related additional energy is small relative to $B_T$ and
cannot compensate for the Coulomb repulsion. Thirdly, the contribution of the Coulomb repulsion 
for $T^{-}$ near the point $\xi \sim 4$ can be evaluated as $<V_c>\sim B_2$ taking into 
account the small binding energy of $T^-$, as well as $B_T/B_2 << 1$. 
The latter means that the $eeh$ system is clustered as $e+(eh)$ and the energy 
separation of the electron from the pair $eh$ is small. According to Eq. (\ref{Sh2_0}) 
one can writes: $B_3=2B_2-<V_c>\approx B_2$ and $<V_c>\sim B_2$. 
 
We have shown that the matrix element $<V_c>$ of the Coulomb interaction for 
two identical particles in the $hhe$ system is 
larger than one in the $eeh$ system. When one disregards the $V_{AA}$ interaction, 
the large MPT in the $hhe$ 
makes the system more bounded compares
with the $eeh$ system, where the contribution of the MPT term is negligible.  
Using scale factor $\alpha \ge 0$ for the ${AA}$ interaction we show  that the 
relation $<V_c(hhe)>\quad \ge \quad <V_c(eeh)>$ is satisfied for increasing $\alpha$  up 
to $\alpha\approx 1$. The system $hhe$ becomes unbound rapidly 
when $\alpha$  is  increasing.
According to Eq. (\ref{Sh2_0}) one can write: $B_3=2B_2+<T_{mp}>-\alpha<V_c>$, 
and can see that $B_3 \approx B_2$ when  the screening is defined by $\alpha\approx 0.7$. 
 The binding energies of the $T^{\pm}$ trions are calculated  
 for different bulk materials based on the Faddeev equation for the $AAB$ 
 system in configuration space. It was found  that the binding 
 energy of $T^{-}$ is relatively small, $B_T/B_2 << 1$, while $T^{+}$ is unbound.  
 The results of the calculations for $B_T$  are consistent with previous 
 computational studies  and are in reasonable agreement with experimental measurements. 
 We explain the origin of the difference of binding energies of $T^{-}$ and $T^{+}$ by using the Schr\"{o}dinger equation written in the system 
 of reference relative to the non-identical particle. There are two terms of the equation 
 which play an important role for the formation of the bound state of a trion when $m_e/m_h<<1$: the Coulomb 
 repulsion between two identical particles and MPT. The MPT, $T_{mp}$, adds a part to the binding 
 energy of the $AAB$ system, while the Coulomb repulsion between $AA$ identical particles decreases the energy. 
 Comparing the bound and  unbound  states of $T^{\pm}$ in considered materials, we show  that 
 hole-hole Coulomb repulsion is stronger in $T^{+}$ than the 
 electron-electron one in $T^{-}$ due to more close localization of the two holes. The last  condition
 is possible due to large contribution of  the MPT.  By introducing the scaling parameter $0\leq \alpha\leq 1$
 and calculation of the binding energy as a function  of  $\alpha$,
 we show that $<V_{AA}>$ is larger for the hole-hole pair.
 
  We illustrated the interplay of these two terms by the hypothetical mass 
 transformation $eeh \to hhe$, which replaces of the masses of the electron and 
 the hole in bulk InN. Using this transformation we demonstrate that the Coulomb 
 repulsion decreases more quickly and the MPT contribution,  $<T_{mp}>$, does not 
 compensate for the binding energy decrease. 
It was demonstrated that $T^{+}$ can be bound by reducing the strength of the Coulomb repulsion 
with the controlled parameter  $\alpha< 1$. 
 
 The properties of the $eeh$ and $hhe$ systems are similar for 3D and 2D models. 
 The similarity is based on the existence of the MPT and the Coulomb repulsion in 
 the Schr\"{o}dinger equation for the both cases. As we show, 
 the interplay of the terms results in the bound or unbound state of the systems. 

This work is supported by 
the National Science Foundation grant Supplement to the NSF grant HRD-1345219 and NASA (NNX09AV07A). 
R.Ya. K. partially supported by MES RK, the grant 3106/GF4.

\end{document}